\documentclass[twocolumn,article,10pt,letter]{IEEEtran}
\usepackage{epsfig,amsfonts}
\usepackage[nolist]{acronym}
\usepackage{graphicx,cite,amssymb,amsmath}
\usepackage{color}
\usepackage{psfrag}
\usepackage{amsbsy}
\IEEEoverridecommandlockouts

\newcommand{\figw}{0.92\columnwidth}

\begin{document}
\begin{acronym}
\acro{LLR}{Log-Likelihood Ratio} \acro{pdf}{probability density
function} \acro{CDF}{cumulative distribution function}
\acro{CCDF}{complementary cumulative distribution function}
\acro{dof}{degrees of freedom}
\end{acronym}

\title{Time Interference Alignment via Delay Offset\\for Long Delay Networks }
\author{Francisco L\'azaro Blasco$^*$, Francesco Rossetto$^*$, Gerhard Bauch$^\dagger$\\
$^*$ Institute of Communications and Navigation\\
DLR (German Aerospace Center), Wessling, Germany 82234\\
$^\dagger$ Department for Communications Engineering\\
Universitaet der Bundeswehr, Munich, Germany\\
Email: {\tt Francisco.LazaroBlasco@dlr.de, Francesco.Rossetto@dlr.de, gerhard.bauch@unibw.de}
\vspace{-24pt}
}

\maketitle
\thispagestyle{empty} \setcounter{page}{0}

\begin{abstract}
\textcolor{black}{Time Interference Alignment is a flavor of
Interference Alignment that increases the network capacity by
suitably staggering the transmission delays of the senders. In this
work the analysis of the existing literature is generalized and the
focus is on the computation of the \acl{dof} for networks with
randomly placed users in a n-dimensional Euclidean
space. In the basic case without coordination among the
transmitters analytical expressions of the sum \ac{dof} can be
derived. If the transmit delays are coordinated, in 20\% of the
cases time Interference Alignment yields additional \ac{dof} with
respect to orthogonal access schemes. The potential capacity
improvements for satellite networks are also investigated.}
\end{abstract}

{\pagestyle{plain} \pagenumbering{arabic}}

\section{Introduction}\label{sec:Intro}

The concept of Interference Alignment (IA) has aroused quite
significant interest in the recent past for its ability to achieve a
first order approximation of the multiuser communication capacity in
some important network configurations~\cite{Cadambe08, Cadambe09a}.
The core idea is to describe the signal as an element of a suitable
space and divide this set into a desired subspace (where the
intended signal should lie) and confine all interference into an
interference subspace. In principle, the desired subspace is
interference free and thus the capacity can grow as the Signal to
Noise Ratio (SNR) increases. It can be shown that the number of
\ac{dof} of the network (by other words, the scaling coefficient of the sum rate)\footnote{\phantom{\tiny{m}}The \ac{dof} for a network and a
certain transmission scheme are defined as: \begin{equation}
 dof=\lim_{P \to +\infty} \frac{C(P)}{\log(P)},
\end{equation}
where $P$ is the transmit power and $C(P)$ is the sum rate of the
network.}
may be significantly beyond 1, i.e., the number of \ac{dof} when the
resources are orthogonally allocated, as for instance in TDMA
\cite{Cadambe08}.

Most studies on IA focus on MIMO based systems, since the
transmitted and received signals are obviously
complex vectors with as many components as the number of antennas
\cite{Gomadan08, Peters09, Schmidt09a, Yetis09, Gollakota09}.
Interference Alignment with MIMO has achieved some degree of
maturity, with results on the capacity and feasibility of
IA~\cite{Schmidt10, Yetis09, Cadambe09a, Cadambe08}, on signal
processing for IA~\cite{Peters09, Gomadan08, Schmidt09a} or even
testbeds~\cite{Gollakota09, ElAyach10}.

\begin{figure*}[!ht]
\begin{center}
\includegraphics[bb = -140 438 605 742,width=1.8\columnwidth,draft=false]{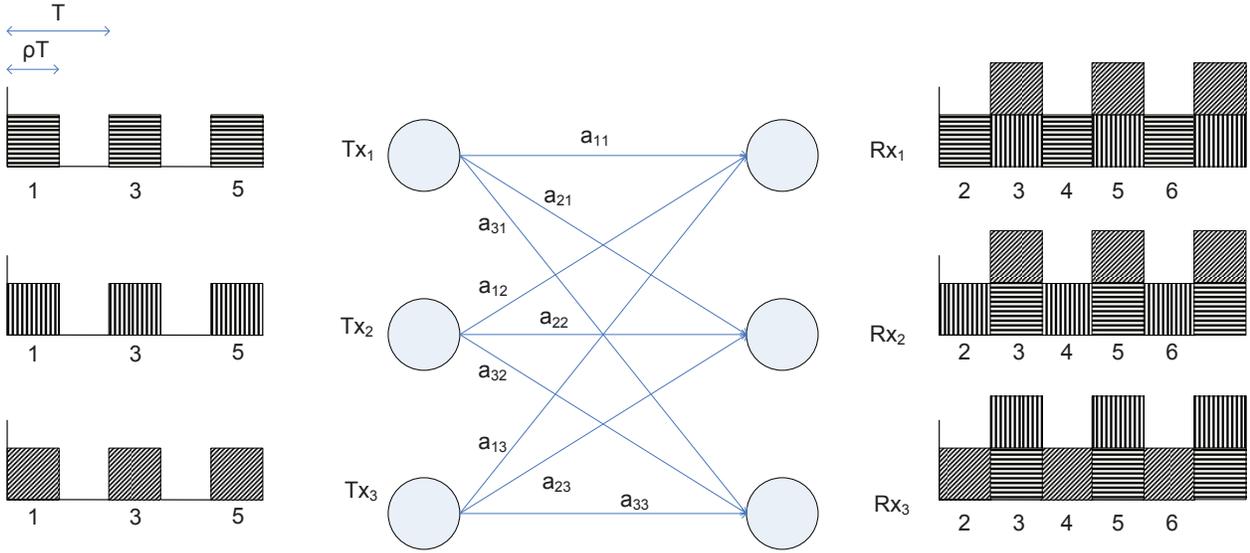}
\centering \caption{The 3-user interference channel. Note that
perfect IA is attained.} \label{fig:IA_time} \vspace{-10pt}
\end{center}
\end{figure*}

Another interesting but not as much studied approach is time based
IA via  delay offset (here named simply time IA for the sake of
brevity)~\cite{Cadambe07, Cadambe08, Mathar09a, Mathar09b, Torbatian10}, where
long propagation delays are exploited. Such scenario can be relevant
for instance in satellite or underwater networks, whose propagation
delays are intrinsically very large. The reference scenario is the
$K$-user interference channel (see Fig.~\ref{fig:IA_time}), where
$K$ transmitters communicate with a dedicated receiver (one per
sender) and all nodes will be assumed to have just one antenna. The
key necessary property for time IA is that the difference of the
propagation delays between transmit-receiver pairs are comparable
to or larger than the packet durations. It was
already noted in the seminal paper~\cite{Cadambe08} that if the
difference between propagation delays is large, it is in principle
possible to perform IA by overlapping the undesired
transmissions, so as to have a non zero amount of time devoid of
interference for the desired signal. Indeed, if $K=3$ and each
sender transmits for $\rho=50\%$ of the time, $K\rho=1.5$ \ac{dof}
are attainable under particular conditions. We shall name "time IA"
this attempt to reduce the portion of time occupied by the
interference through mutual coordination of the transmitters.

It is arguable that when time IA is meaningful, this method may show
some advantages with respect to other types of IA, like MIMO IA.
For instance, the transmitters need to know the propagation delays
rather than the complex gain of a fading channel, and normally the
former can be estimated quite accurately  for instance by means of a
GPS receiver and may not be so time varying as a fading channel.
Moreover, propagation delay variations can be partly predicted if
information on the user speed is available, which is not possible
for many other forms of IA. Hence, it may be speculated, although
yet to prove, that time IA is more robust than other implementations
of IA.

By perfect IA it is meant that at every receiver the total time
where interference is present is equal to the duration of a single
transmission (see Fig.~\ref{fig:IA_time} for an example). The work
of~\cite{Mathar09b, Mathar09a} has shown that in $\mathbb{R}^n$ it
is possible to deploy $n+1$ pairs which achieve perfect time IA and
each couple transmits $\rho=50\%$ of the time, if transmitters are
synchronized and start transmitting at the same time instant. The
sum rate will then scale with $(n+1)\rho\log_2(1+\Lambda)$, where
$\Lambda$ is the SNR (assumed to be equal for all pairs). Note that
the capacity of orthogonal access schemes would grow with
$\log_2(1+\Lambda)$. Although it is not known how many pairs can be
placed in $\mathbb{R}^n$ if the transmitters are allowed to delay
the beginning of their transmission, it is to be expected that more
than $(n+1)$ can be placed \cite{Mathar09a}. For example in
\cite{Cadambe07} it is shown how to place $4$ pairs in $2$
dimensions. Thus, the sum rate gain can be particularly large but it
requires very special and regular placements of the transmitters and
receivers. In real world networks, the positions of the users cannot
be arbitrarily decided and hence it is not always possible to reduce
the span of time taken by interference to its minimum value of
$\rho$ at every receiver. However, a question of practical relevance
is to investigate how much capacity can be attained by suitably
aligning the transmissions of the terminals in a network with long
delays. While it will be very unlikely to approach as many as
$(n+1)\rho$ \ac{dof}, it is important to understand whether more
than one degree of freedom can be obtained with non negligible
probability.

The contribution of this work lies in an analysis of
the achievable \ac{dof} with time IA in an $n$-dimensional Euclidean
space with random user positions. Section~\ref{sec:sys_model}
introduces the system model and Section~\ref{sec:non_cooperative}
investigates the attainable \ac{dof} without optimizing the
transmission delay, where insightful analytical results can be
derived. Section~\ref{sec:cooperative} studies the sum rate when the
transmission delays are jointly optimized among the senders, and in
particular the performance improvements due to this coordination are
highlighted. Section~\ref{sec:application} investigates the amount
of additional \ac{dof} of time IA for a satellite network and
finally Section~\ref{sec:Conclusions} draws the conclusions.

\section{System Model}\label{sec:sys_model}
We will denote matrices by capital letters. $m_{i,j}$ will be the
element at the $i$-th row and  $j$th column of a matrix $M$.
$\mathbb{R}^n$ shall stand for the the n-dimensional Euclidean
space. Further  $M \mod{p}$ will denote the elementwise modulo $p$
operator of matrix $M$. The rectangle function or unit pulse
centered around $0$ and of length $1$ will be expressed as $rect(x)$
and $\delta(x)$ will be used to denote Dirac's delta.

The $K$-user interference channel will be considered, with $K=3$, as
shown in Fig.~\ref{fig:IA_time}, in which $K$ transmitters
communicate with $K$ receivers. We shall assume that transmitter 1
wants to communicate with receiver 1, transmitter 2 with receiver 2
and so on. Let us consider the general case in which transmitters
and receivers are placed in $\mathbb{R}^n$ and the delay between two
nodes is proportional to their Euclidean distance. Note that $n$ is
generic. The propagation delay among all the nodes in the channel
can be expressed in a $(K \times K)$ matrix $A$, where $a_{i,j}$ is
the propagation delay between transmitter $j$ and receiver $i$. Time
will be divided into slots of length $T$. Transmitters will be
allowed to transmit only for a time $\rho T$ in every time slot,
$\frac{1}{K}<\rho\leq \frac{1}{2}$. The ratio of the transmit
duration over the length of a time slot, $\rho$, will be called duty
cycle. This framework is more general than the cases considered in
literature up to now \cite{Cadambe08,Mathar09a,Mathar09b}, but it
also includes the canonical example of interference alignment by
means of delay offsets, where transmitters are allowed to transmit
for a fraction $\rho = \frac{1}{2}$ of the time. If we would choose
$\rho<\frac{1}{K}$ an orthogonal resource allocation in time or
frequency would outperform interference alignment. Moreover, there
is no reason for choosing $\rho>1/2$, since it is known that it is
not possible to obtain more than 1/2 \ac{dof} per user
\cite{Cadambe08}.

The elements $a_{i,j}$ shall be assumed to take values between $(0,
\gamma)$ where $\gamma>>T$. Given the fact that  the time allocation
scheme used is periodic with period $T$, it is useful to define:
\begin{equation}
 B=\frac{A\mod{T}}{T}.
\end{equation}

$B$ will be referred to as the normalized
propagation delay matrix. The elements $b_{i,j}$ take values in $[0,
1)$. Additionally transmitters will be allowed to delay the start of
their transmission. We will denote by $\delta_i$ the initial
transmission delay of transmitter $i$. Let us define a new matrix,
\begin{equation}
D =\frac{(A+\hat{\Delta})\mod T}{T},
\end{equation}
where $\hat{\Delta}$ is a matrix whose elements,
$\hat{\delta}_{i,j}$, correspond to the initial transmit delay $\delta_j$ of
transmitter $j$, $\hat{\delta}_{i,j} = \delta_j $, $\forall i$. The
matrix  $D$ together with $\rho$  completely specify the IA network.

\section{Analysis for non-coordinated transmitters}\label{sec:non_cooperative}
In this section the case in which the transmit delays are not
coordinated is considered. Hence transmitters do not have any
knowledge of $A$, $B$ or $D$. The initial transmission delay,
$\Delta_i$, will be assumed to be independent and uniformly
distributed between $0$ and $T$. While this transmission scheme does
not attempt to perform interference alignment, analytical
expressions can be derived which actually correctly predict some
behaviors of a network whose transmit delays are coordinated to
approximate interference alignment. We shall start with a normalized
propagation delay matrix $B$ whose elements are independently picked
from a uniform distribution between $0$ and $1$. This may seem
unrealistic at first. However, this assumption is true if we assume
transmitters and receivers to be placed randomly in a "big enough"
section of $\mathbb{R}^n$. Under these hypotheses, it is
straightforward to infer that the elements in $D$ are also
independent and uniformly distributed between $0$ and $1$. Let us
also recall that we work under the assumption that $1/3 \leq \rho
\leq 1/2$.

Let us denote by $\alpha_i$ the \ac{dof} attained by pair $i$, and
as $\phi = \sum_i{\alpha_i}$ the sum of the \ac{dof} achieved by the
3 transmitter receiver pairs. In this section we derive analytically
the \ac{pdf} of the sum \ac{dof}, $f(\phi)$, of the 3-user
interference channel. Because of the symmetry of the problem,
$f(\alpha_i)$ does not depend on $i$ and will be just denoted as
$f(\alpha)$. Let us say, for example, concentrate on receiver $1$.
The \ac{dof} $\alpha$ for this pair correspond with the ratio of
time in which the signal from transmitter $1$ is received at
receiver $1$ without any interference. For simplicity it will be
assumed that the interference from transmitter $2$ starts at $t=0$,
$d_{1,2} = 0$. Our first step to calculate $f(\alpha)$ will be
calculating the \ac{CDF} of $\alpha$ conditioned to $d_{3,1}$:
\begin{equation}
\mathbf{F}(\alpha | d_{3,1}) = P(\boldsymbol{\alpha} \leq  \alpha | d_{3,1}).
\end{equation}

\begin{figure}[t]
\begin{center}
\includegraphics[bb = 42 60 578 781,width=0.74\columnwidth,draft=false]{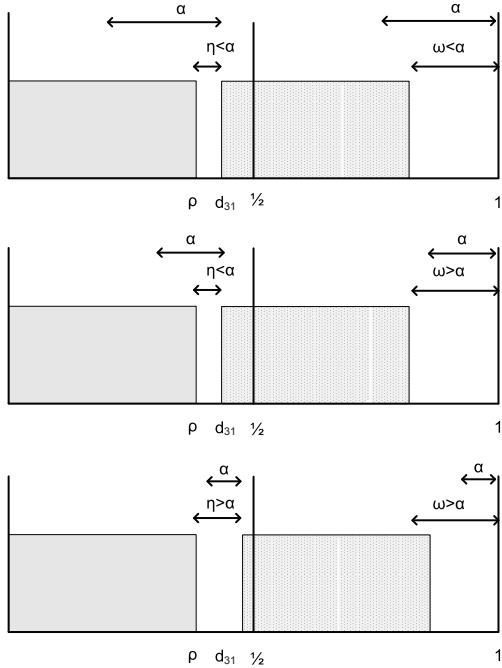}
\centering \caption{The three possible configurations of the intervals to calculate $f(\alpha)$. The upper part corresponds to $\eta<\alpha$ and  $\omega<\alpha$. The middle part to $\eta<\alpha$ and  $\omega>\alpha$ and the bottom part
to $\eta>\alpha$ and  $\omega>\alpha$}
\label{fig:alpha}
\vspace{-10pt}
\end{center}
\end{figure}

Let us first consider the case $d_{3,1}<\frac{1}{2}$.  For
simplicity two auxiliary variables $\eta$ and $\omega$ will be
introduced. $\eta$  and $\omega$ are associated with
the interference free time before and after $d_{3,1}$, respectively. A negative
value implies that no interference free time exists. Fig.~\ref{fig:alpha} shows the three different cases which have to be
taken into account depending on the value of $\eta$  and $\omega$.

The first case is depicted in the upper part of Fig.~\ref{fig:alpha}. In this case both $\eta$ and $\omega$ are smaller
than $\alpha$. These conditions translate respectively into $\rho +
\alpha
> d_{3,1}$ and $ d_{3,1} + \rho + \alpha >1$. Both
conditions can be put together as $\rho + \alpha > d_{3,1} > 1 -
\rho - \alpha$. Therefore, this situation can only take place if
$\rho + \alpha > 1 - \rho - \alpha$, which implies $\alpha >
\frac{1}{2} - \rho$. In this case since $\eta$ and $\omega$ are
smaller than $\alpha$ no matter which values $d_{1,1}$ takes it is
impossible to achieve $\alpha$ \ac{dof}:
\begin{equation}
\mathbf{F}(\alpha | d_{3,1}) = 1, \alpha > \frac{1}{2}-\rho.
\label{eq:first_case}
\end{equation}

In the second case $\eta<\alpha$ and  $\omega>\alpha$. This
situation is shown in the middle part of Fig.~\ref{fig:alpha}. These
conditions imply $\rho + \alpha < d_{3,1}$ and $ d_{3,1} + \rho +
\alpha <1$. Or equivalently: $\rho + \alpha < d_{3,1} > 1 - \rho -
\alpha$. Given the values of $\rho$ considered, this second case is
always possible independently from the value of $\alpha$. In this
case the probability of achieving at least $\alpha$ \ac{dof} is not
zero. Recalling the assumption that $d_{i,i}$ are independent and
uniformly distributed from $0$ to $1$:
\begin{equation}
\begin{array} {ll}
\mathbf{F}(\alpha | d_{3,1}) &= P(\boldsymbol{\alpha} \leq  \alpha | d_{3,1})= \\
&= P(d_{1,1} < d_{3,1} + \alpha ) + P(d_{1,1} > 1- \alpha )= \\
&= d_{3,1} + \alpha + \alpha =  d_{3,1} + 2 \alpha.
\label{eq:second_case}
\end{array}
\end{equation}

The last case, shown in the bottom part of Fig.~\ref{fig:alpha}
corresponds to the situation in which $\eta>\alpha$ and
$\omega>\alpha$. This corresponds to $\rho + \alpha > d_{3,1} > 1 -
\rho - \alpha$. Which can only happen if  $\alpha < \frac{1}{2} -
\rho$. For this case the \ac{CDF} of $\alpha$ conditioned to $d_{3,1}$
is:
\begin{equation}
\begin{array} {ll}
& \mathbf{F}(\alpha | d_{3,1}) = P(\boldsymbol{\alpha} \leq  \alpha | d_{3,1})= \\
&= P(d_{1,1} < \alpha ) + P(d_{3,1}-\rho-\alpha <d_{1,1} < d_{3,1}-\rho ) + \\
& + P(d_{3,1}<d_{1,1} < d_{3,1}+ \alpha ) + P(d_{1,1} > 1- \alpha ) = \\
 &= 4 \alpha.
 \label{eq:third_case}
\end{array}
\end{equation}

So far we have assumed $d_{3,1}<\frac{1}{2}$. It is easy to see that
considering the case $d_{3,1}>\frac{1}{2}$ is equivalent to
exchanging $\eta$ and $\omega$. By virtue of this symmetry, the
\ac{CDF} of $\alpha$ can be calculated as:

\begin{equation}
\begin{array} {ll}
\mathbf{F}(\alpha )&= \int_0^1 \mathbf{F}(\alpha | d_{3,1}) f(d_{3,1}) \, \mathrm{d} d_{3,1} = \\
&= 2 \int_0^{\frac{1}{2}} \mathbf{F}(\alpha | d_{3,1}) f(d_{3,1}) \, \mathrm{d} d_{3,1}\\
\label{eq:int_0}
\end{array}
\end{equation}

In order to solve the integral in \eqref{eq:int_0} two different
cases must be considered, $\alpha < \frac{1}{2} - \rho$ and $\alpha
> \frac{1}{2} - \rho$. If $\alpha < \frac{1}{2} - \rho$, $\mathbf{F}(\alpha | d_{3,1})$
is given by \eqref{eq:second_case} and \eqref{eq:third_case}:
\begin{equation}
\begin{array} {ll}
&\int_0^{\frac{1}{2}} \mathbf{F}(\alpha | d_{3,1}, \alpha < \frac{1}{2}-\rho) f(d_{3,1}) \, \mathrm{d} d_{3,1} = \\
& = \int_0^{\rho+\alpha} (d_{3,1} + 2 \alpha) f(d_{3,1}) \, \mathrm{d} d_{3,1}+ \int_{\rho+\alpha}^{\frac{1}{2}} (4 \alpha) f(d_{3,1}) \, \mathrm{d} d_{3,1} =\\
& = -\frac{3}{2} \alpha^2 -  \alpha \rho + 2 \alpha + \frac{3}{2} \rho^2 , \, \alpha \leq \frac{1}{2}-\rho
\label{eq:int_2}
\end{array}
\end{equation}

Similarly if $\alpha > \frac{1}{2} - \rho$, $\mathbf{F}(\alpha |
d_{3,1})$ is given by \eqref{eq:first_case} and
\eqref{eq:second_case}:
\begin{equation}
\begin{array} {ll}
&\int_0^{\frac{1}{2}} \mathbf{F}(\alpha | d_{3,1}, \alpha > \frac{1}{2}-\rho) f(d_{3,1}) \, \mathrm{d} d_{3,1} = \\
& = \int_0^{1-\rho-\alpha} (d_{3,1} + 2 \alpha) f(d_{3,1}) \, \mathrm{d} d_{3,1}+\\
& + \int_{1-\rho-\alpha}^{\frac{1}{2}} 1 f(d_{3,1}) \, \mathrm{d} d_{3,1} =\\
& = -\frac{3}{2} \alpha^2 -  \alpha \rho + 2 \alpha + \frac{3}{2} \rho^2, \, \alpha \leq \frac{1}{2}-\rho
\label{eq:int_3}
\end{array}
\end{equation}
Note that \eqref{eq:int_2} and \eqref{eq:int_3} yield the same
result. From   \eqref{eq:int_0}, \eqref{eq:int_2} and
\eqref{eq:int_3} the \ac{CDF} of $\alpha$ is obtained:
\begin{equation}
\begin{array} {ll}
\mathbf{F}(\alpha )= -3 \alpha^2 - 2 \alpha \rho + 4 \alpha + \rho^2,  \, \alpha \leq \frac{1}{2}-\rho
\label{eq:cdf}
\end{array}
\end{equation}

By taking the first derivative of \eqref{eq:cdf} with respect to
$\alpha$ it is possible to obtain the pdf of $\alpha$:
\begin{equation}
\begin{array}{ll}
f(\alpha) &= \frac{\partial \mathbf{F}(\alpha)}{\partial \alpha} = (4 - 2 \rho - 6 \alpha) rect \left( \frac{\alpha - \frac{\rho}{2}}{\rho}\right) +\\
&+ \rho^2 \delta(\alpha) +  (1-4 \rho + 4 \rho^2) \delta(\alpha-\rho)
\label{eq:pdf_alpha}
\end{array}
\end{equation}

Now that the \ac{pdf} of the \ac{dof} for \emph{one} user is known,
the \ac{pdf} of the \ac{dof} for $3$ users can be calculated as the
convolution of $3$ single user \acp{pdf}. After some straightforward
but tedious computations the \ac{pdf} of $\phi$ is obtained:
\begin{equation}
\begin{array}{ll}
f(\phi) &= a^3 \delta(\phi) + 3a^2b \delta(\phi-\rho) + 3ab^2 \delta(\phi-2\rho)+ \\
& + b^3 \delta(\phi-3\rho) + p_1(\phi) rect \left( \frac{\phi - \frac{\rho}{2}}{\rho}\right)  + \\
& + p_2(\phi) rect \left( \frac{\phi - \frac{3 \rho}{2}}{\rho}\right) + p_3(\phi) rect \left( \frac{\phi - \frac{5 \rho}{2}}{\rho}\right),
\label{eq:pdf_phi}
\end{array}
\end{equation}
where:
\begin{equation}
\begin{array}{ll}
a&= \rho^2 \\
b&= 1 - 4  \rho + 4\rho^2\\
p_1(\phi) &= - 6p^5 - 6p^4\phi + 12p^4 + 32p^3\phi^2 - 48p^3\phi + 6p^2\phi^3 +\\
& - 48p^2\phi^2 + 48p^2\phi - 9p\phi^4 + 48p\phi^3 - 48p\phi^2 - \frac{9}{5}\phi^5 +\\
& + 18\phi^4 - 48\phi^3 + 32\phi^2 \\
p_2(\phi) &=\frac{303}{5}p^5 + 39p^4\phi - 198p^4 - 118p^3\phi^2 + 240p^3\phi + \\
& +78p^3 - 12p^2\phi^3 + 204p^2\phi^2 - 426p^2\phi + 96p^2 + \\
& +18p\phi^4  - 96p\phi^3 + 78p\phi^2 + 96p\phi - 48p + \frac{18}{5}\phi^5 + \\
& - 36\phi^4 + 114\phi^3  - 136\phi^2 + 48\phi\\
p_3(\phi) &=- \frac{273}{5}p^5 - 33p^4\phi + 186p^4 + 86p^3\phi^2 - 192p^3\phi +\\
& - 42p^3 + 6p^2\phi^3 - 156p^2\phi^2 + 378p^2\phi - 168p^2 + \\
& - 9p\phi^4 + 48p\phi^3 - 30p\phi^2 - 96p\phi + 78p - \frac{9}{5}\phi^5 +\\
& + 18\phi^4 - 66\phi^3 + 104\phi^2 - 66\phi + 12
\label{eq:p}
\end{array}
\end{equation}

\begin{figure}[ht]
\begin{center}
\includegraphics[width=\figw,draft=false]{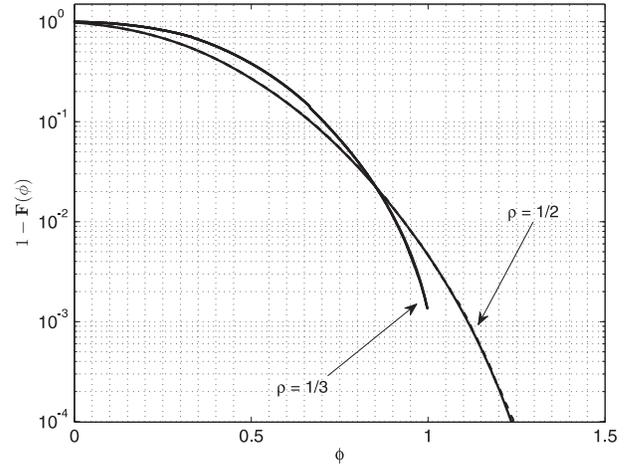}
\centering \caption{\ac{CCDF} of $\phi$ for $\rho=0.5$ , and $\rho=\frac{1}{3}$ for non-coordinated transmitters. Theoretical results
are plot with  discontinuous lines and simulation results with solid lines.}
\label{fig:noncooperative}
\vspace{-10pt}
\end{center}
\end{figure}

Fig.~\ref{fig:noncooperative} shows the \ac{CCDF} of $\phi$ for
$\rho = 0.5$ and $\rho = \frac{1}{3}$. The figure shows results of
numerical simulations and analytical formulas. It can be observed
how the analytical results and simulation results match perfectly.
Note that the dirac deltas in \eqref{eq:pdf_phi} should create
discontinuities in the \ac{CCDF} of $\phi$. For $\rho=\frac{1}{2}$
the effect of the deltas can not be appreciated because they
have a very low weight. However for
$\rho=\frac{1}{3}$ there is a discontinuity at $\phi=3\rho=1$. It
can be observed how at $\phi=3\rho=1$ the \ac{CCDF} goes from
 $b^3= 1.4E-3$ to zero.

\begin{figure}[ht]
\begin{center}
\includegraphics[width=\figw,draft=false]{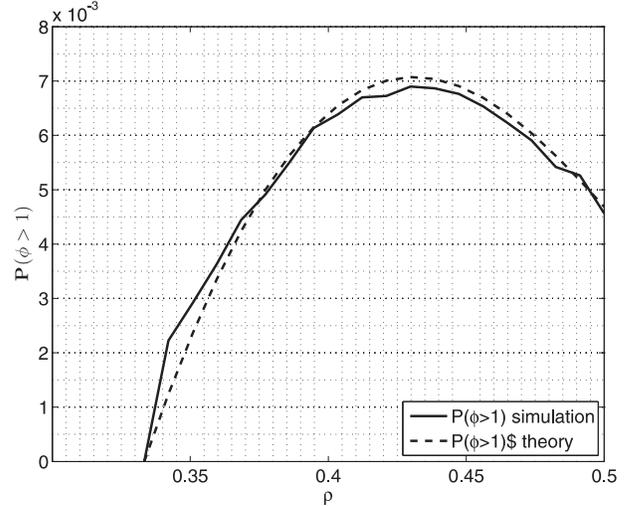}
\centering \caption{Probability that the sum of \ac{dof} in the network exceeds 1 against the duty cycle $\rho$ for non-coordinated transmitters. Theoretical results are plotted with a discontinuous line and simulation results with a solid line.}
\label{fig:phi_bigger_1}
\vspace{-10pt}
\end{center}
\end{figure}

With the results provided in  \eqref{eq:pdf_phi} it is possible to
calculate which is the value of $\rho$ that maximizes the
probability of exceeding some given number of \ac{dof}. For example,
the probability of exceeding 1 \ac{dof} can be calculated as:
\begin{equation}
\begin{array}{ll}
P(\phi>1) = \int_1^{3\rho}f(\phi) \, \mathrm{d} \phi
\label{eq:opt_rho}
\end{array}
\end{equation}
After some elementary calculus it can be found that the value of
$\rho$ which maximizes $P(\phi>1)$ is $\rho_{opt}= 0.4305$.
Fig.~\ref{fig:phi_bigger_1} shows $P(\phi>1)$ as a function of
$\rho$. In the figure numerical and theoretical results are shown.
It can be observed how $P(\phi>1)$ has  maximum at $\rho_{opt}$.
Moreover simulation results match quite well theoretical formulas.
Note also that the probability that the 3-user interference network
without cooperation exceeds 1 \ac{dof} is in any case very low,
around $7E-3$. If transmitters do not coordinate at all, it is very
unlikely to achieve a high capacity.

\section{Numerical results for coordinated transmitters}\label{sec:cooperative}

Section~\ref{sec:non_cooperative} characterized the performance of a
transmission scheme whose transmitters had no knowledge of the
matrices $A$ and $B$ and did not coordinate their transmit delays
$\delta_i$. However, according to the interference alignment
paradigm, the transmitters should arrange their signals so as to
minimize the impact of the interference they generate.  In this
section all senders are assumed to have perfect knowledge of
matrices $A$ and $B$. The target metric has been the maximization of
the sum \ac{dof} $\phi$, which can be achieved by two goals: first,
the reduction of the amount of time that is occupied by the
interference at each receiver and secondly to minimize the overlap
between the desired signal and the interference. Hence, the senders
arrange their transmission delays $\delta_i$ in order to maximize
$\phi$. This optimization is non trivial generally speaking and it
has been performed numerically in a centralized fashion. Note that
the purpose of our work is not to develop a distributed algorithm
for this task but rather to obtain insight into the potential
capacity gains attainable by time IA. Moreover, the impact of the
duty cycle $\rho$ has been studied as well. Indeed, a larger $\rho$
will increase both the duration of the useful signal and also of the
mutual interference, hence it is non trivial to infer whether the
duty cycle should approach its maximum possible value $1/2$ or it
should be close to the minimum $1/3$.

\begin{figure}[ht]
\begin{center}
\includegraphics[width=\figw,draft=false]{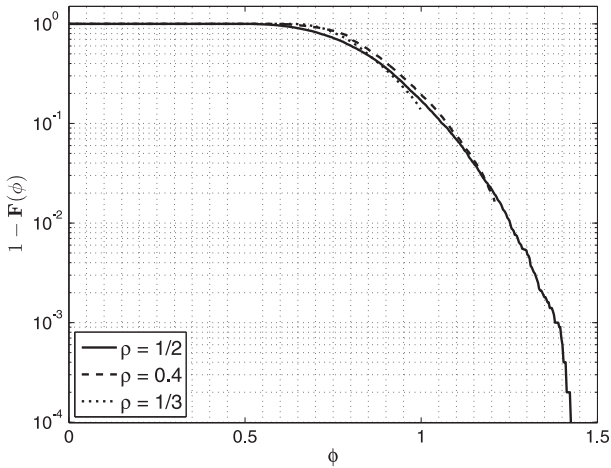}
\centering \caption{\ac{CCDF} of $\phi$ for coordinated transmitters for $\rho=0.5$ , $\rho=0.4$ and $\rho=\frac{1}{3}$. Note the scale is same as in Fig.~\ref{fig:noncooperative}}
\label{fig:CCDF_phi_opt}
\vspace{-10pt}
\end{center}
\end{figure}

\begin{figure}[ht]
\begin{center}
\includegraphics[width=\figw,draft=false]{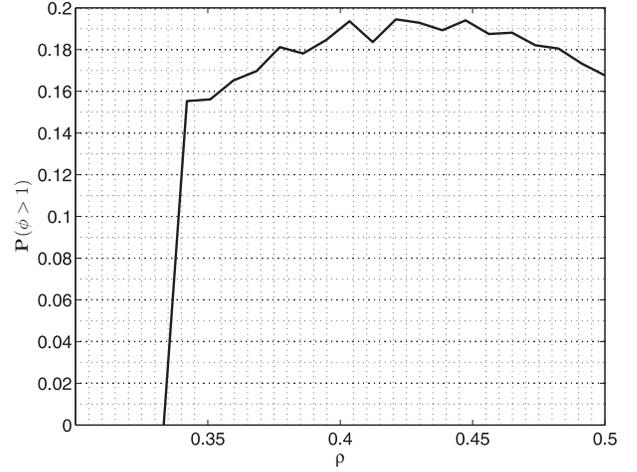}
\centering \caption{Probability that the sum of \ac{dof} in the network exceeds 1 against the duty cycle $\rho$ for coordinated transmitters.}
\label{fig:phi_bigger_1_opt}
\vspace{-10pt}
\end{center}
\end{figure}

Fig.~\ref{fig:CCDF_phi_opt} shows the CCDF of $\phi$ for
$\rho\in\{1/3,\,2/5,\,1/2\}$. As it can be noticed, the optimization
of the transmit delays has brought a significant capacity
improvement. For instance, the 90-th percentile for $\rho=1/2$ has
increased from $\phi=0.65$ to $1.05$. Moreover, while the
uncoordinated case would have a non negligible probability of
attaining almost no  \ac{dof} (see Fig.~\ref{fig:noncooperative}),
in the coordinated case the probability density function would be
non zero only after 0.5. Another important observation is that the
duty cycle plays a role especially for the right tail of the CCDF.
First of all, the maximum number of \ac{dof} is upper bounded by
$K\rho=3\rho$, and this can be seen in Fig.~\ref{fig:CCDF_phi_opt},
since the CCDF does not exceed that limit. Furthermore, the value of
$\rho$ does impact the probability of attaining a high capacity,
because for instance the curve for $\rho=2/5$ achieves better
performance than $\rho=1/3$ and $\rho=1/2$ for $0.5<\phi<1.1$.
Indeed, Fig.~\ref{fig:phi_bigger_1_opt} depicts the probability of
obtaining more than one degree of freedom, and is the equivalent of
Fig.~\ref{fig:phi_bigger_1} with transmit delay optimization. One
can notice that apart for the minimum value $\rho=1/3$, the
coordination of the transmit delays attains more \ac{dof} than an
orthogonal access scheme with probability 15\%. This probability is
further increased to 19.5\% for the optimum value of
$\rho\simeq0.42$ and thus this metric has increased by over one
order of magnitude with respect to the non coordinated case.
Moreover, note that the optimum value of $\rho$ for the
uncoordinated and coordinated case look very similar (0.43 and 0.42
respectively), which suggests that $\rho\simeq0.43$ is a good duty
cycle for possibly many configurations. It may be expected that
further optimizations are possible with respect to these quite
simple strategies and therefore further capacity improvements may be
attainable.

\section{Applications for satellite networks}\label{sec:application}

As stated in the introduction, one of the motivating scenarios for
our study are satellite networks.  These systems are distinguished
by long delays and it is of interest to investigate whether time IA
may bring an advantage for instance for multisatellite networks.
Many reasons can be brought to have multiple satellites
in the same frequency, but two are of particular relevance. First of
all, the system may reuse the bandwidth very aggressively by
deploying multiple satellites. Moreover, it may be more viable from
an economic point of view to first deploy one satellite and launch
additional systems when the revenues pick up. Hence, a first order
investigation of how much capacity can be achieved by time IA can
have practical importance. It should also be noticed that transmit
delay coordination is in fact already in place in some TDMA
standards for the return link, like DVB-RCS, which require that the
signals of the users arrive at specific time instants at the
satellite. These delay corrections are computed by the network
gateway and therefore from a system point of view there would be
almost no cost in implementing time IA for multiple satellites
directed by a single gateway: this controller should compute in any
case the appropriate transmit delays of the ground users, hence the
difference at the gateway lies mainly in the
software algorithm that computes such numbers, not in the hardware.

The analyzed scenario is still the $K$ network, where $K$ satellites
communicate with some stations on ground in the same frequency and
with partial overlap of time. In fact, this approach works equally
well in both transmission directions (ground to space and viceversa)
and can be applied without major differences in the forward and
return link. In our study, the focus is on geostationary satellites, because a
large number of communication satellites are moving along this
orbit. The performance of time IA depends quite heavily on the
distances between the satellites. Indeed, from the previous
discussion, the larger the difference between the
transmitter-receiver distances, the better our previous analysis
applies. Moreover, also better performance can be in general
attained, as it will be shown. However, large distances between the
satellites imply rather different orbital positions. Therefore the
minimal orbital separation present today for geostationary
satellites is assumed, which corresponds to $0.5^\circ$. Three
satellites at 24.5, 25 and 25.5 degrees east have been assumed. The
constellation is positioned over Europe, and thus the ground
stations are generated randomly but their latitude and longitude is
constrained in the ranges $[35^\circ,\,55^\circ]$ north and
$[-10^\circ,\,20^\circ]$ east, respectively. A duty cycle
$\rho=0.43$ has been adopted, since it was optimal for the previous
two configurations and finally the delays are jointly optimized.

\begin{figure}[ht]
\begin{center}
\includegraphics[width=\figw,draft=false]{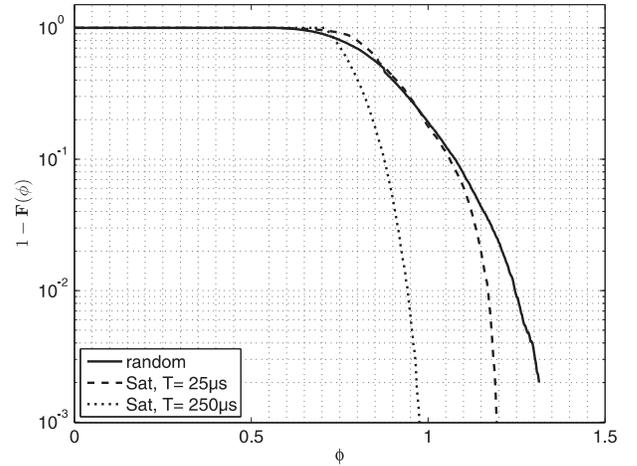}
\centering \caption{\ac{CCDF} of $\phi$ for cooperative transmitters satellite. $\rho=0.43$ is assumed.}
\label{fig:CCDF_phi_opt_sat.eps}
\vspace{-10pt}
\end{center}
\end{figure}

An important element here is the duration of $T$. According to our
previous discussion, the shorter $T$, the more random the matrix
looks like. We have evaluated some possible values of $T$ in the
range from tens to hundreds of microseconds, which imply bandwidth
of tens to hundreds kHz and are still reasonable for satellite
systems. Indeed, the forward link bandwidth is often in the order of
5 MHz, so these $T$ can work for short bursts of symbols.
Fig.~\ref{fig:CCDF_phi_opt_sat.eps} shows the CCDF of the sum
\ac{dof} $\phi$ for this configuration with $T=25$ and $250$ $\mu$s.
In addition, the curve for completely random delay matrices is also
depicted. It can be noticed that long $T$ like $250\mu$s impose a
heavy loss in terms of \ac{dof}, since the delay matrices $B$ do not
look random anymore but rather the delays become quite similar. For
shorter $T$ in the order of $25\mu$s, the CCDF significantly
deviates only at the tail, thus randomness of the delay matrix is
important especially to achieve the configurations with large number
of \ac{dof}.This deviation comes from the fact that the position of
the $3$ satellites is not random but always fixed. Note that with
$T=25\mu$s, the CCDF evaluated at  $\phi=1$ is $0.2$, that is to say
that the probability of outperforming TDMA is 20\%, as in the
previous setting. It should be noticed that in these scenarios, the
number of ground users can be extremely large (it could easily be
several thousands), and thus a smart scheduler could group the users
in configurations that do achieve the high end of the CCDF curve.
Thus, by virtue of multiuser diversity (which is truly abundant in
this setting), the high performance of time IA might be routinely
attained.

It is possible to relax the constraints on $T$ by increasing the
separation between satellites. However, some satellite antennas can
be quite large (in the order of 1 m of aperture) and thus their
mainlobe would be narrow. In this case, no mutual interference would
be present by design, since the narrow lobes would suppress the
other interference. Hence, satellites in far orbital slots would be
more suitable for mobile receivers, whose antennas must be small and
thus would suffer from inter-satellite interference. A
multi-satellite constellation for mobile users has actually been
rolled out in the XM-Sirius satellite radio system in order to
provide satellite diversity, therefore such systems are in fact
already reality and therefore the proposed scheme may be practically
relevant.

\section{Conclusions}\label{sec:Conclusions}

This work has studied the capacity improvements brought by time
interference alignment in a more general setting than its
predecessors~\cite{Cadambe07, Cadambe08, Mathar09a, Mathar09b}. In particular,
the possibility to approximate perfect interference alignment by
means of delay and duty cycle optimization has been investigated and
a first-order evaluation to the satellite case has been performed as
well. Analytical expressions have been derived for the  \ac{dof} of
the non coordinated case, which yielded useful predictions for a
system whose transmitters coordinate to approach
interference alignment. The results show that by simple ideas there
can be a non negligible capacity improvement and further
optimizations are currently under investigation.

\bibliographystyle{IEEEtran}
\bibliography{IEEEabrv,studio3}

\end{document}